\documentclass{iau} 
\usepackage{graphicx}

\def\kms{km~s$^{-1}$}
\def\lcdm{$\Lambda$CDM}
\def\msun{M$_{\odot}$}

\title[What Telescopes Will Discover]
{The Future of Dwarf Galaxy Research: \\ What Telescopes Will Discover}

\author[Martha P. Haynes]  
{Martha P. Haynes$^1$}

\affiliation{$^1$Cornell Center for Astrophysics and Planetary Science, Cornell University,\\
530 Space Sciences Building, 122 Sciences Drive, Ithaca, NY 14853, USA \\ email: {\tt haynes@astro.cornell.edu}
}

\pubyear{2018}
\volume{344}  
\jname{Dwarf Galaxies from the Deep Universe to the Present}
\editors{K.B.W. McQuinn and S. Stierwalt eds.}

\begin{document}

\maketitle

\begin{abstract}

Dwarf galaxies make ideal laboratories to test galaxy evolution paradigms and cosmological models. Detailed studies of dwarfs across the spectrum allow us to gauge the efficacy of astrophysical processes at play in the lowest mass halos such as gas accretion, feedback, turbulence and chemical enrichment. Future observational studies will deliver unprecedented insights on the orbits of dwarf companions around the Milky Way, on their star formation histories and on the 3-D internal motions of their stars. Over large volumes, we will assess the impact of local environment on baryon cycling and star formation laws, leading to a full picture of the evolution of dwarfs across cosmic time. In combination, future discoveries promise to trace the history of assembly within the Local Group and beyond, probe how stars form under pristine conditions, and test models of structure formation on small scales.

\end{abstract}

\firstsection 

\section{Introduction}

The low mass, low luminosity dwarfs are the most common galaxies in the universe 
but, because of their diminutive characteristics, are hard to detect except at 
nearby distances. Despite their numerical dominance, dwarf galaxies do not contribute 
much to luminosity or mass functions but they are exceptional in other ways. Most of 
them are dark matter (DM) dominated, many by a large factor. Both their stars and 
interstellar media have low metallicities, comparable to the abundances expected of 
the early universe. Since hierarchical models suggest that low mass halos are the 
fundamental building blocks of massive galaxies, dwarf galaxies may serve as local 
analogs of the earliest galaxies and thus reflect the physical conditions of early 
star formation (SF). Likewise, because of their low mass, they are expected to be 
much more strongly affected both by internal feedback processes and by external 
environmental influences. Thus dwarf galaxies serve as excellent cosmological 
probes and laboratories for galaxy evolution studies.  

Numerous published reviews of the state of dwarf galaxy research contain more 
through and elaborate detail than presented here. 
Among the most important reviews focusing
on the Local Group (LG) dwarf population are
those of \cite[Hodge (1971)]{Hodge71}, \cite[Mateo (1998)]{Mateo98} and 
\cite[McConnachie (2012)]{McConnachie12}). Driven by the discovery of their dominance by
number in the Virgo cluster (\cite[Binggeli, Sandage \& Tammann 1985]{Binggeli85}),
the early type dwarf population was
reviewed by \cite[Ferguson \& Binggeli (1994)]{Ferguson94}.
\cite[Tolstoy, Hill \& Tosi (2009)]{Tolstoy09}
presented a excellent summary of the star formation histories (SFHs), chemical
abundances and kinematics
of LG dwarfs while most recently, \cite[Bullock \& Boylan-Kolchin (2017)]{Bullock17} discuss 
how the number, distribution and structural properties of dwarf galaxies challenge current cosmological models.

Many papers in this volume contribute significant
insight into particular directions and are more focused than this review.
An excellent overview of dwarf galaxies is presented
by Andrew Cole in this volume and Laura Sales discusses the
successes and challenges of numerical simulations. Following on
those works, in this review I briefly discuss the state
of dwarf galaxy observations in the context 
of the future research directions where observations play a key
role in understanding. I apologize to those whose exciting work I 
am unable to call out because of page limitations; this was an
amazing and stimulating conference.

\section{The Dwarf Galaxy Population}

The classification of a galaxy as a ``dwarf'' has historically revolved
around its small optical luminosity, size and, in most cases,
surface brightness. In his review, \cite[Hodge (1971)]{Hodge71} set
the maximum luminosity (M $>$ -15) of a dwarf one magnitude fainter than the
Small Magellanic Cloud (SMC; M $\sim$ -16). More recently
\cite[Bullock \& Boylan-Kolchin (2017)]{Bullock17} have categorized a
galaxy as a ``dwarf'' if it has a stellar mass M$_{*}$ $<$ 10$^9$
\msun. They further subdivide the dwarf class into ``bright'' dwarfs with
10$^7$ $<$ M$_{*}$ $<$ 10$^9$ \msun, ``classical'' dwarfs with
10$^5$ $<$ M$_{*}$ $<$ 10$^7$ \msun, and ``ultra-faint'' dwarfs (UFDs;
\cite[Willman et al. 2005]{Willman05}) with
10$^2$ $<$ M$_{*}$ $<$ 10$^5$ \msun. 

Many other authors have used other descriptors
to subcategorize dwarfs according to distinguishing characteristics such as
morphology: 
dwarf ellipticals (dE), dwarf spheroidals (dSph), dwarf irregulars (dIrr
or dI). Although in most categorizations the Magellanic Clouds (MCs) are
brighter than dwarfs, galaxies comparable especially to the SMC are sometimes
referred to as Magellanic type dwarfs (dIm). The ``transition'' dwarfs (dTrans)
show a mixture of early- and late-type dwarf characteristics, and may be
in an intermediate stage of transitioning between types.
Some dwarf ellipticals are nucleated (dE,N) while others are not.
The ultra compact dwarfs (UCD), (\cite[Hilker et al. 1999]{Hilker99};\cite
[Drinkwater et al. 2000]{Drinkwater00}) are extremely small, with half-light
radii of $<$ 100 pc, as compact as globular clusters, but
with typical masses greater than $10^6$ 
\msun. UCDs have been found in many clusters, from Virgo 
(\cite[Zhang et al. 2015]{Zhang15}) to
distant ones, and
also in groups and around relatively isolated spirals. Blue compact dwarfs (BCD) are 
small low luminosity, 
star-bursting systems characterized by low metallicity and ofter showing
evidence suggestive of interactions. Prototypical examples are IZw18
(\cite[Searle \& Sargent 1972]{Searle72}, \cite[Lebouteiller et al. 2017]
{Lebouteiller17})
and SBS0335 (\cite[Izotov et al. 1990]{Isotov99}). 
Tidal dwarf galaxies (TDG) are systems formed out of the debris of
galaxy encounters such as the Antennae (
\cite[Mirabel, Dottori \& Lutz 1992]{Mirabel92}). TDGs are distinguished
among the dwarf population by their lack of dark matter (\cite[Lelli et al. (2015)]{Lelli15})
and greater metallicity for their luminosity (\cite[Duc et al. 2007]{Duc07};
\cite[Lee-Waddell et al. 2018]{Lee-Waddell18}), matching
better the expectations of their brighter parent galaxies.
As discussed by Kristine Spekkens 
in this volume, recent attention has been focused
on galaxies of extreme low surface brightness, the ultra diffuse galaxies, UDGs
(\cite[van Dokkum et al. 2015]{vanDokkum15}). UDGs are very extended
with effective radii comparable to that of the MW but with stellar masses of $\sim$100 
times less. While the majority of the UDGs are
found in clusters, others appear to be relatively isolated 
(\cite[Martinez-Delgado et al. 2016]{MartinezDelgado16},
\cite[Greco et al. 2018]{Greco18}). Some actually contain
abundant supplies of neutral gas (\cite[Leisman et al. 2017]{Leisman17}).

Noted also by many previous authors, \cite[Tolstoy, Hill \& Tosi (2009)]{Tolstoy09}
point out in their Figure 1 that the dSphs, dIrrs and the star-bursting BCDs 
trace similar relations in the luminosity-size and luminosity-surface brightness
planes. In fact, their updated Figure 1 reaffirms the conjecture by 
\cite[Kormendy (1985)]{Kormendy85} that dwarfs fall along the same relations
defined by the more luminous galaxies
without discontinuity. 
While the UFDs are clearly separated in luminosity, they show similar trends
to those of the brighter dwarfs, suggesting some commonality.
Various arguments propose that outliers result from 
environmental processes, particularly those which reduce the size and/or
luminosity of the original galaxy.

\subsection{Dwarfs and their Environment}

The close relationship between dwarf galaxy morphology and local environment is long
known, even before the discovery of the UFDs. Most of the dwarfs in the 
LG lie within the virial radius of the Milky Way (MW) or Messier 31 (M31; Andromeda), 
but most of the gas-rich and currently star-forming ones are found
beyond 300 kpc from the host (\cite[Tolstoy, Hill \& Tosi 2009]{Tolstoy09},
\cite[McConnachie 2012]{McConacchie}). The observed MW morphological segregation
suggests strong links between the local environment and the processes which
shut down and/or trigger SF episodes. 

While the environments of the LG and nearby groups are relatively benign,
rich clusters offer a glimpse of dwarf evolution under extreme conditions
of interaction. Deep surveys of clusters are revealing increasing populations of
diverse dwarf galaxies, extending across the full range of dwarf sub-classes.
For example, the similarity in compactness of UCDs to globular clusters (GCs) hints
of a common linkage, that they may form by mergers of large numbers of
GCs. Alternatively, UCDs may be the central nuclei of normal dwarfs stripped by 
tidal interactions in the rich cluster environment (\cite[Zhang et al. 2015]{Zhang15}.

\section{Dwarf Galaxies as Cosmological Probes}
In their recent review of ``dwarf problems'' with \lcdm, \cite[Bullock \& Boylan-Kolchin (2017)]{Bullock17} 
discuss numerous critical areas where dwarf galaxies raise challenges for our current
cosmological models. These have been to a large extent reviewed by Laura Sales in her plenary talk
included in this volume. Here, I review the observational constraints on the number, structure and distribution
of dwarf galaxies which future observations will address.

\subsection{The Census of Dwarf Galaxies} 
Although \lcdm ~simulations predict the existence of thousands of
low mass halos with masses large enough to support the molecular cooling 
needed to form stars, there is a strong mismatch between the number of 
dark matter (DM) subhalos predicted by the simulations and the number of satellites
of the Milky Way that have been observed (\cite[e.g. Klypin et al. 1999]{Klypin99}). 
As mentioned above, the pace of discovery of faint galaxies in the LG has advanced
rapidly with the deployment of facilities that offer both sensitivity and
wide area coverage. The discoveries of new UFD MW satellites by the Sloan Digital
Sky Survey (SDSS) prompted the reexamination of the predictions of the MW
satellite luminosity function with a better understanding of completeness limits
and the biases introduced by sky coverage and Galactic extinction (\cite[Tollerud et al.
2008]{Tollerud08}). In the intervening years, advances in the predictions
of simulations and observations have diminished but not yet fully eliminated
the mismatch. As noted by Andrew Cole in his talk included in this volume,
20 new LG dwarfs have been discovered in the last year, bringing the census
of galaxies closer than 1.5 Mpc to 107, including more than 50 MW satellites
and more than 35 around M31.

\lcdm ~predicts than DM halos should be filled with smaller 
substructures and thus that the satellites of the MW should have 
satellites of their own. \cite[Sales et al. (2013)]{Sales13}
have predicted based on semi-analytic models of galaxy formation that
analogs of the LMC should host one SMC mass satellite as well as
between 5 and 40 satellites of mass within 1/1000th of the LMC.
The recent report by 
\cite[Kallivayalil et al. (2018)]{Kallivayalil18} that four of the recently
discovered satellites in the general direction of the MCs are in fact associated
with the Magellanic system is confirmation that ``satellites of satellites''
exist. The continued study of these objects will be crucial to our understanding of
the relationship between stellar mass and halo mass at the lowest masses: how
low mass halos can hang onto their baryons in the presence of both internal
feedback and environmental processes.

Blind HI 21 cm line surveys offer an alternate path toward the discovery of 
gas-bearing dwarf galaxies. The recently-completed ALFALFA extragalactic HI survey
(\cite[Haynes et al. 2018]{Haynes18}) covered 7000 sq. deg. of high latitude sky
out to c$z \sim$ 18000 \kms, detecting thousands of galaxies with baryonic masses
less than 10$^9$ \msun, and more than 100 with HI masses less than 3 $\times 10^7$ \msun.
As discussed separately by Betsey Adams and Julio Navarro in this volume and
\cite[Giovanelli et al. (2010)]{Giovanelli10}, it is possible that some low
velocity ALFALFA HI detections may be very low mass LG subhalos with few, or no, stars.

A challenge to the identification of gas-bearing but optically-dark dwarfs is 
that their HI line signals overlap in velocity with galactic and circumgalactic 
ISM phenomena, particularly the HI high velocity clouds (HVCs). Furthermore,
there is no way to determine the distance to the galaxy by the HI line alone.
\cite[Adams, Giovanelli \& Haynes (2014)]{Adams14} presented a first catalog of
candidate objects, identified within the ALFALFA survey database 
by relatively small size and isolation  as Ultra Compact High Velocity Clouds,
(UCHVCs).  As summarized by Adams in this volume, on-going work aims to determine
the distances to several other UCHVCs identified in
ALFALFA (\cite[Janesh et al. 2017]{Janesh17}). 

\begin{figure}[t!]
\centering
\includegraphics[height=2.15in]{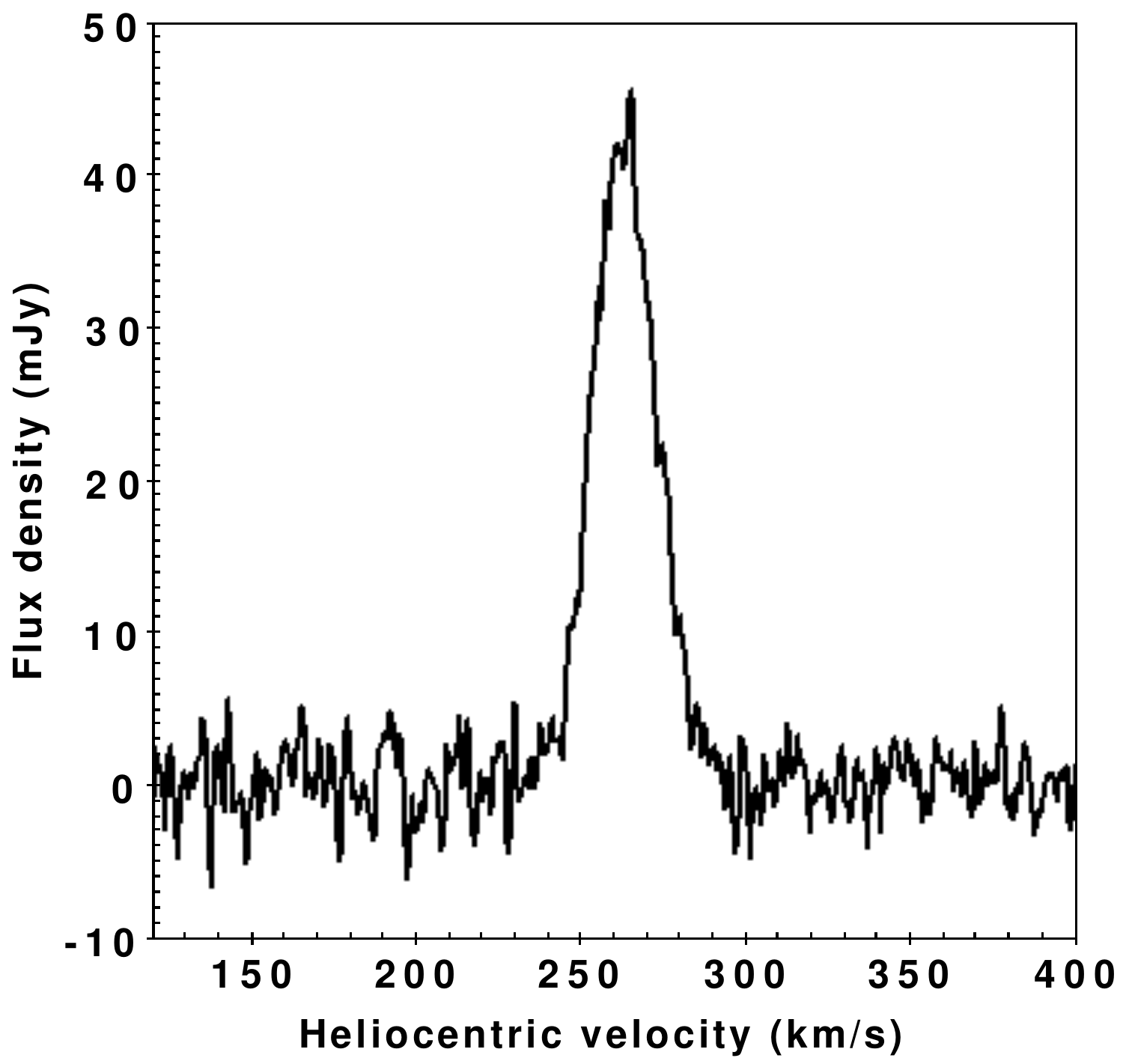}
\includegraphics[height=2.23in]{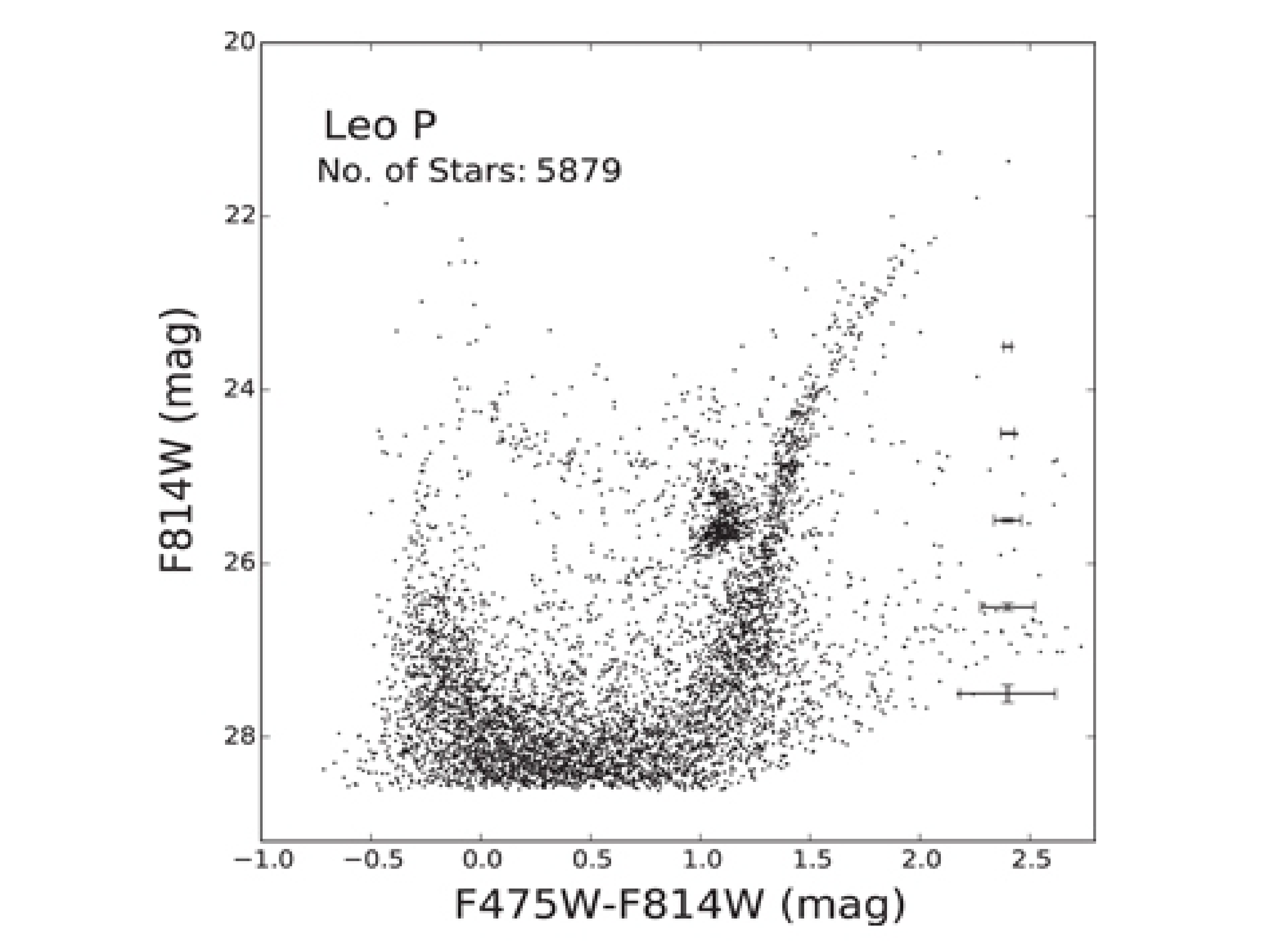}
\vspace{-0.1cm}
\caption{Left: High resolution spectrum of the global HI line emission obtained 
with the L-band wide receiver at Arecibo. Right: Color-magnitude diagram of Leo P 
from HST ACS imaging included as Figure 2 of \cite[McQuinn et al. (2015)]{McQuinn15}.}
\label{fig:figLeoP}
\end{figure}

The nearby dwarf Leo P is a prototype of a nearby dwarf galaxy discovered by its HI
line emission as detected by the ALFALFA survey.
While the HI emission from Leo P meets the criteria of an UCHVC, its
optical counterpart was noticed immediately 
(\cite[Giovanelli et al. 2013]{Giovanelli13}) because it has a single HII region 
(\cite[Rhode et al. 2013]{Rhode13}) evident in public imaging datasets.
Figure \ref{fig:figLeoP} shows a high resolution HI line global profile (obtained with
the Arecibo single beam, higher gain L-band wide receiver system)
and the color magnitude diagram based on HST imaging used to determine 
the distance, confirming its location on the outskirts
of the LG. The HI line signal is strong and corresponds to
8.1 $\times$ 10$^5$ \msun ~at the distance of 1.62 $\pm$ 0.15 Mpc
(\cite[McQuinn et al. 2015]{McQuinn15}).
The spectrum shown in Figure\ref{fig:figLeoP} has a resolution of 
about 1.4 km/s; the best fit Gaussian has a full width at half 
maximum of 22.4 $\pm$ 0.3 \kms;
\cite[Bernstein-Cooper et al. (2014)]{BernsteinCooper14} derive a rotational
velocity of 15 $\pm$ 5 \kms ~from the VLA velocity field. Uncertainties remain
in the inclination of the disk, the contribution of turbulence to the
observed line width, and the relationship of the velocity traced by the HI
to the halo circular velocity. But the fact remains that Leo P has a very low
SF rate despite the fact that its baryonic mass is dominated by HI.

Based on its CMD (left panel of Figure \ref{fig:figLeoP}), 
\cite[McQuinn et al. (2015)]{McQuinn15} find that Leo P has formed stars at a 
relatively constant rate over its lifetime and
that it may be what a low mass dSph would look
like today if it managed, because of its relative isolation, to hold onto its HI gas.

\subsection{The Structure of Dwarf Galaxy Halos}

\lcdm ~simulations with dark matter only predict that halos should have density profiles
that rise steeply at small radius $\rho \propto r^{-\gamma}$ with $\gamma$ in the range
(0.8,1.4) on the scale of dwarf galaxies
(\cite[Navarro et al. 2010]{Navarro10}). However, as inferred from the observational
surface brightness and kinematic data, the central regions of dark-matter dominated
dwarfs seem less dense and less cuspy than predicted by the simulations. Deriving the
density profiles at very small radii is difficult, complicated by resolution and
seeing issues as well as geometry. For example, \cite[Genina et al. (2018)]{Genina18})
point out that the distribution of stellar populations within dwarfs can be significantly
elongated. Such departures from sphericity can hide cuspiness and even suggest 
a core when there is a cusp. The addition of baryonic processes
such as feedback into the simulations similarly can modify their predictions so that a
discrepancy between observations and predictions is not yet proved.

\subsection{Scaling Relations for Dwarf Galaxies}

\begin{figure}[t]
\begin{center}
 \includegraphics[width=5in]{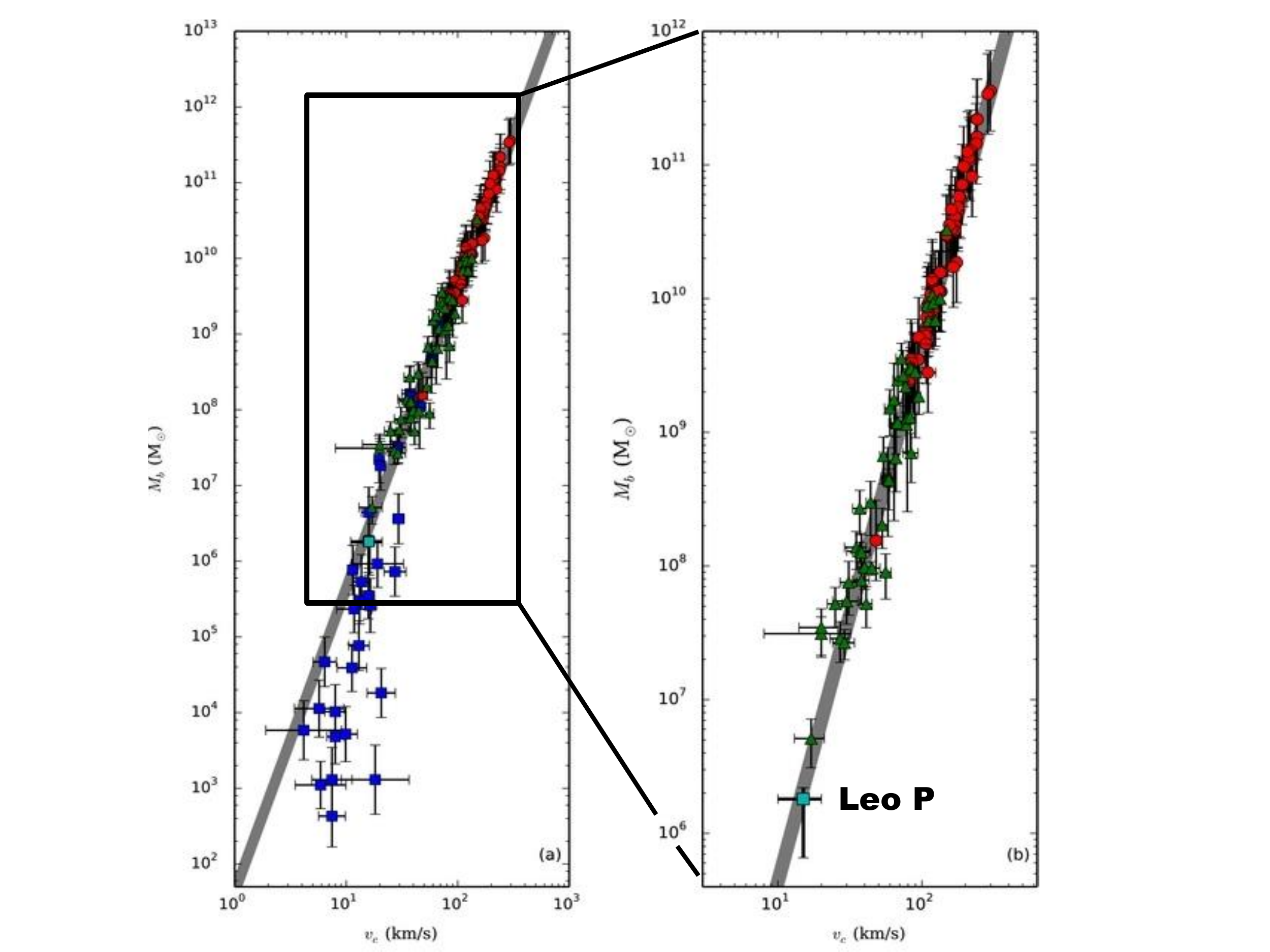} 
 \caption{The baryonic Tully-Fisher relation for local galaxies, adapted 
 from \cite[Bernstein-Cooper et al. (2014)]{BernsteinCooper14}. The dataset 
 is reproduced from \cite[McGaugh (2012)]{McGaugh12}. The left panel shows 
 the full range of the observational dataset, including the LG dSphs, while 
 the right panel covers only the range explored by the gas-dominated galaxies. 
 The gray shaded line shows the best fit to the gas-dominated galaxies as 
 derived in \cite[McGaugh (2012)]{McGaugh12}. The labeled point in the 
 right panel shows the location of the dwarf irregular galaxy Leo P, 
 discovered by its HI line emission in the ALFALFA survey.}
   \label{fig:btfr}
\end{center}
\end{figure}

As discussed in some detail by \cite[Bullock \& Boylan-Kolchin (2017)]{Bullock17},
it is quite surprising that the baryonic components of galaxies
seem so tightly related to the dynamical properties inferred for their host halos,
given the very wide range of their morphologies, star formation histories, dark-to-luminous
matter fractions and locations within larger scale structures. Figure \ref{fig:btfr},
adapted from \cite[Bernstein-Cooper et al. (2014)]{BernsteinCooper14} shows one
example of such scaling relations, the baryonic Tully-Fisher relation (BTFR:
\cite[McGaugh et al. 2000]{McGaugh00}). Based
mainly on data compiled by \cite[McGaugh (2012)]{McGaugh12}, the right panel
of Figure \ref{fig:btfr}
shows that the BTFR holds over 6 orders of magnitude to at least M$_{baryon}$ $\sim$
10$^6$ \msun. At baryonic masses below that, the UFDs deviate
from the relation, possibly due to the effects of their local environment. 

An extension of the BTFR known as the radial acceleration relation (RAR) has been
noted by \cite[McGaugh, Lelli \& Schombert (2016)]{McGaugh16}. As discussed in
greater depth by those authors and by 
\cite[Bullock \& Boylan-Kolchin (2017)]{Bullock17}, real galaxies show a much
wider diversity in rotation curve shape at fixed velocity
compared to what is predicted by
the simulations. A tight coupling of the total baryon content (gas plus stars)
with the halo mass introduces lots of astrophysical complications whose impact
on the interpretation of the observational data need
to be understood (\cite[Verbeke et al. 2017]{Verbeke17}).
    
\subsection{Satellite Planes}

\lcdm ~simulations do not predict the existence of large departures
from isotropy in the distribution of satellites nor departures from
random motions. However, in
1976, Lynden-Bell and, independently, Kunkel and Demers 
(\cite[Lynden-Bell 1976]{LyndenBell76}) noticed the possible association
of distant globular clusters and dwarf galaxies; their proposed planes
were offset by 30$^\circ$ with Lynden-Bell's overlapping the recently
discovered Magellanic Stream. 
With the discovery of additional dwarfs and the measurement of 
more precise distances,
better definition of the Milky Way satellite plane was made by
\cite[Kroupa, Theis \& Boily (2005)]{Kroupa05} and
\cite[Metz, Kroupa \& Jerjen(2007)]{Metz07}. Now dubbed the
``Vast POlar Structure'' (VPOS),
\cite[Pawlowski, McGaugh \& Jerjen (2015)]{Pawlowski15}
have shown that the system forms a flattened body with half-thickenss
$\sim15$ kpc and radius $\sim$40 kpc. Furthermore, 
a similar preferential arrangement has been identified around
M~31 (\cite[McConnachie \& Irwin 2006]{McConnachie06}, \cite[Ibata et al. 2013]{Ibata13})
Canonical thought has suggested that these flattened structures
might be explained by the preferential accretion of satellites along
filaments. However, work presented at this conference by 
Oliver M\"{u}ller (\cite[M\"{u}ller et al. 2018]
{Muller18}) has shown not only that a plane of satellites exists around 
the nearby galaxy Centaurus A,
but also that the statistical analysis of their observed kinematics 
suggests that the satellites are co-rotating, a feature not predicted by 
current simulations. Better distances and proper motions of more MW and M31
satellites and studies of dwarfs in more distant systems will be needed to
understand the occurrence of such anisotropic structures and their origin.

\section{Dwarf Galaxies as Probes of Galaxy Evolution}

Because of their proximity, the stellar constituents of LG dwarf galaxies can
be studied in great detail both through color-magnitude diagrams and spectroscopic 
studies of resolved stars. With increasing resolution, the atomic and molecular 
constituents of nearby dwarfs offer clearer insight into the interstellar media in
low mass, low metallicity systems. Together the body of evidence enables the
exploration of
how dwarfs acquire, enrich and retain their gas while building up
their stellar mass.

\subsection{Star Formation Histories}

The availability of detailed 
color-magnitude diagrams (CMDs) for growing numbers of LG dwarfs
reinforces the concept of morphological segregation in terms of 
star formation histories, particularly
among the dwarfs which are not satellites of the MW or M31. 
As shown by, among others, \cite[Tolstoy, Hill \& Tosi (2009)]{Tolstoy09}, 
\cite[Weisz et al. (2014)]{Weisz14} and
\cite[Gallart et al. (2015)]{Gallart15},
dwarfs in the LG show a remarkable variety of SFHs. Whereas the
more isolated dwarfs show a wide variation in their SFHs, the CMDs of
closest-in MW UFDs all suggest that SF
ceased 10 Gyr ago. More distant dSph satellites show evidence
of young and intermediate age populations. 
\cite[Skillman et al. (2017)]{Skillman17} find that a small sample
of M31 dSphs show a significant range of quenching times but none
that could be considered recent ($<$ 5 Gyr). For a larger sample with
somewhat shallower photometric data, \cite[Martin et al. (2017)]{Martin17} also
found that a large fraction of the M31 dwarf galaxies have extended SFHs
inconsistent with early star formation episodes that were rapidly shutdown. 
Further studies will
extend the depth of the CMDs in larger samples, allowing inferences
on the importance of quenching at reionization and possible 
relationships to orbital parameters and location relative to 
satellite planes.

An impressive clue to the evolution of dwarf galaxies is contained in
the luminosity-metallicity relation or the alternative version,
the stellar mass-metallicity relation. Combining stellar abundances
[Fe/H] from \cite[Kirby et al. (2013)]{Kirby13} with nebular ones [O/H] from 
\cite[Berg et al. (2012)]{Berg12}, Figures 2 and 3 of
\cite[Hidalgo (2017)]{Hidalgo17} demonstrate the similarity in the
two versions as measured by the two techniques. It is quite amazing that
dwarfs with the same mass (or luminosity) but very different SFHs can apparently
end up with the same metallicity. Such an outcome may arise because
the least massive systems lose so many of their heavy elements 
during bursts that SF has only a slow impact on their buildup. 

Locally, spectroscopic studies of individual resolved stars provide a detailed
look at the chemical enrichment history in dwarfs, complementary to their SFHs.
Sylvia Ekstr\"{o}m, in this volume, reviewed the importance of metallicity in
the role that massive stars play in chemical enrichment. As discussed by
\cite[Tolstoy, Hill \& Tosi (2009)]{Tolstoy09}, because stars of different
masses contribute different elements on different timescales, studies of
the pattern of abundance variations of individual resolved stars can reveal 
how chemical enrichment has taken place during the galaxy's SFH. At low metallicity,
the abundance ratios of heavy elements are expected to deviate from those
of normal populations (\cite[Emerick et al. 2018]{Emerick18}). 
Indeed, extremely-metal poor stars in the MW, believed to
be among the oldest population, are strongly overabundant in carbon and often
also in Ni, O and Mg. Detailed studies of the abundance of various
chemical species in individual stars in UFDs offer evidence 
that those objects may in fact represent very low mass halos whose stars
were formed at very early times (\cite[Spite et al. 2018]{Spite18}).
Future high sensitivity, high spatial and
spectral resolution studies will make use of the wealth of information
conveyed by studies of abundances and their patterns.

\subsection{The Interstellar Medium in Dwarf Galaxies}

Observations of the interstellar medium (ISM) in dwarf galaxies offer insights 
into the processes that lead to the conversion of gas into stars in 
low metallicity environments as well
as those that lead to the shut-down of SF due to feedback, stripping, etc.
In this volume, Suzanne Madden presents an overview of the ISM properties 
of dwarfs, looking in particular at what observations of
interstellar gas and dust tell us about how the gas phases operate and what 
processes control dust and gas evolution. The low metallicity 
of low mass dwarfs bears importantly on cooling rates and therefore 
on dust and molecule formation. Among other effects, the grain size 
distribution, composition and structure are encoded on the efficiency
of photoelectric heating. The dust-to-gas ratio (D/G) shows 
a roughly linear relation with metallicity at higher values of Z  
but at Z/Z$_{\odot}$ $<$ 0.1, D/G plummets steeply
(\cite[R\'{e}my-Ruyer et al. 2015]{RemyRuer15}). 
In spirals, H$_2$ formation on the surfaces of grains is quite 
efficient but in dwarfs, the timescale for formation
becomes very long, exceeding a Hubble time. Although the process is
not fully understood, it seems that 
metals are not incorporated into dust at low metallicity in the same 
way as at higher Z, with the result that low metallicity galaxies require
more time to accumulate heavy elements for efficient grain growth.

Alberto Bolatto, in this volume, has reviewed the status
of our understanding of the formation and cooling of molecular gas
and its relationship to SF, particularly in the Magellanic Clouds.
The cold ISM is notoriously difficult to detect in dwarf systems, but
recent CO detections (e.g., \cite[Elmegreen et al. 2013]{Elmegreen13}, 
\cite[Shi et al. 2016]{Shi16}) have shown that such gas does exist
in their low metallicity environments. However, at low metallicity, 
CO becomes a poor
tracer of H$_2$. \cite[Bolatto, Wolfire \& Leroy (2013)]{Bolatto13} 
have shown that the H$_2$ to CO conversion factor blows up at low metallicities.
H$_2$ can exist outside of the denser regions where the CO is
dissociated but the H$_2$ remains self-shielded. This distinction makes
the denser CO core considerably smaller and thus makes much of the
molecular mass ``CO-dark''. In low metallicity dwarfs, the CO dark gas may
harbor the bulk of the H$_2$ reservoir. 

While there is molecular gas in low metallicity dwarfs, it is accompanied apparently 
by a decreased SFR per gas mass. Of particular relevance to probe SF in pristine conditions such as found in the early universe,
future studies will need to understand the role of HI vs H$_2$ in how 
dwarfs form stars and the impact of their often-bursty SFHs.

\subsection{The Regulation of Star Formation in Dwarfs}

The fraction of gas in cold phases depends on the processes
governing the collapse of cold gas (what triggers star formation) 
and on the ones that heat it up, thereby preventing collapse (what 
keeps the star formation rate so low?). The 
star formation laws at work in the Milky Way and
spirals in general do not seem to
apply in low mass, low surface brightness, low metallicity dwarfs
(e.g., \cite[Kennicutt 1998]{Kennicutt98}, 
\cite[Kennicutt \& Evans 2012]{Kennicutt12}). 
Recent works by Shi and colleagues, summarized most recently in
\cite[Shi et al. (2018)]{Shi18}, have
used spatially resolved measurements of the SFR as
well as stellar and gas masses to explore the deviations seen
in low surface brightness galaxies and the outer disks
of spirals and dwarfs from
the usual relationship between the SFR surface density and the 
local gas surface density
${{\rm{\Sigma }}}_{\mathrm{SFR}}$ $\propto$ ${{\rm{\Sigma }}}_{\mathrm{gas}}^{N}$
with $N$ $\sim$ 1.4,
the ``Kennicutt-Schmidt Law''. Rather, they propose an ``extended Schmidt law'',
${{\rm{\Sigma }}}_{\mathrm{SFR}}$ $\propto$ ${{\rm{\Sigma }}}_{\mathrm{gas}}
{{\rm{\Sigma }}}_{\mathrm{*}}^{N}$,
which includes an additional dependence on the stellar mass surface density
arising from the role the gravity of the stellar population
plays in setting the pressure of the ISM. Those authors further
point out that, under similar conditions, SF was inefficient in the
early universe. Using aperture synthesis HI maps from the
Faint Irregular Galaxy GMRT Survey (FIGGS),
\cite[Roychowdhury, Chengalur \& Shi (2017)]{Roychowdhury17} have
shown that this extended law provides a much better fit
than the simpler relation.

Additionally, it is well understood that feedback
associated with the evolution of massive stars play an important
role in regulating star formation in fragile dwarf disks. A number of
talks at this conference discussed recent results on how various
feedback mechanisms can influence low mass galaxies. As reviewed
by Samantha Penny in this volume, recent observations
conducted as part of the SDSS-IV MaNGA survey have presented evidence
for AGN feedback in galaxies just at the top end of the dwarf class
considered here (M$_*$ $\sim$ 10$^9$ \msun, \cite[Penny et al. 2018]{Penny18}).
Volker Heesen discussed using radio continuum emission as an extinction
free tracer to probe the influence of stellar feedback in dwarfs
using IC 10 as an example (\cite[Heesen et al. 2018]{Heesen18}).
Vianney Lebouteiller presented results showing that the heating of the neutral
gas in the ever-enigmatic IZw~18 mainly results from photoioniozation
by radiation from a bright X-ray binary in that galaxy
(\cite[Lebouteiller et al. 2017]{Lebouteiller18}). Conference organizer
Kristy McQuinn presented a poster on winds in starbursting low
mass galaxies following on previous work showing that the massive star
wind timescales are comparable
to the starburst duration(\cite[McQuinn et al. 2018]{McQuinn18}). 
As a result, the gas is driven to larger distances and 
less material is recycled than might have been expected. 

Future studies combining
multiwavelength tracers will provide a holistic understanding of
how different internal and external factors influence the mechanisms
that trigger, sustain and shut down star formation in dwarfs. The importance
of environment will require studies not just of galaxies close to their
hosts or in clusters, but also ones in voids 
(\cite[Makarov et al. 2017]{Makarov17}) or in pairs of isolated
dwarfs (\cite[Privon et al. 2017]{Privon17}).

\section{Conclusion: The Future of Dwarf Galaxy Research}

In the preceding sections, I have tried to call out some of the
most intriguing recent results related to observations of dwarf
galaxies and what questions they raise.  Here I summarize some of 
questions that future observations are likely to answer or at least
provide significant insight into their answers.

{\bf How many dwarf galaxies are there?} \hskip 5pt
To appreciate the rapid rate of dwarf discovery, it
should be noted that the majority of LG dwarfs identified
today are fainter than any galaxy known in 2000 and, as Andrew
Cole noted in his overview talk, that
more than 20 new LG dwarfs have been found in the last year. 
Following on this tremendous pace of discovery, future wide area and
highly sensitive surveys at optical wavelengths including the on-going
Dark Energy Survey, the Hyper Suprime-Cam survey and ultimately,
the Large Synoptic Survey Telescope
will provide a truly robust census of the LG down to the minimum halo mass 
expected to host a bona fide galaxy. For example, \cite[Newton et al. (2018)]{Newton18}
have assessed the current limitations in observational
constraints on the number of dwarf MW satellites and make
predictions for future surveys. Their analysis estimates that
there should be 124 $(+40,-27)$ (68\%, statistical error) dwarfs
brighter than M$_V \sim$ 0 within R $<$ 300 kpc of the MW. And, they
conjecture that the LSST will be able to detect half of them,
despite the large swath of sky hampered by galactic
extinction and sky coverage limitations.

In parallel with the optical surveys, the suite of 21 cm wide area mapping
programs conducted with Arecibo/
ALPACA, FAST, WSRT/APERTIF, MeerKAT and ASKAP will probe
the low mass end of the HI mass function to find any gas-bearing halos even 
if they do not have associated stars. Some of the latter may be the 
reionization-limited HI-bearing low mass haloes dubbed RELHICS 
(\cite[Ben\'{i}tez-Llambay et al 2017]{Benitez17}) and discussed 
by Julio Navarro in this volume. Not only will the next generation
of HI surveys discover many additional dwarfs,
but the resolved HI surveys with interferometers will provide
further insight into how the observed disk rotation reflects the
halo circular velocities (\cite[Papastergis \& Shankar 2016]
{Papastergis16}). Furthermore, they will explore the lowest HI mass
population in hundreds to thousands of dwarfs in nearby groups beyond the LG.

{\bf How do satellites exist within halos?} \hskip 5pt
A main objective of the Gaia mission was to unravel the assembly 
history of the MW. Its recent second data release has
shown its success in enabling the analysis of the spatial
distribution, kinematics, age and abundance of stars across the
MW. Future studies will aim to link together the clues to assembly
history in order to construct a self-consistent time sequence of what
happened when, why and how. 
Initial works based on Gaia have already suggested the association
of MW components with satellite accretion events (\cite[Helmi et al. 2018]{Helmi18}).
Among the objectives of future programs to identify larger and larger
samples of dwarfs and to study their individual stars and stellar 
populations will be to trace, with the obvious distance-dependence
of detail, the assembly history of dwarfs throughout and 
beyond the LG. 
As discussed in this volume by Isabel Santos-Santos, interpretation
of the current observational data on satellite planes
remains a challenge, particularly because
of the relatively small numbers of dwarfs used in defining planes and limitations
in sky coverage particularly due to Galactic extinction. Future
deeper and wider observational studies of satellites not just in the Local
Group but beyond will be required to test the advances in understanding made
in the simulations.

{\bf How does star formation in low metallicity gas proceed?} \hskip 5pt
Theoretical predictions of the internal structure of ISM clouds in dwarfs 
suggest that in low metallicity environments there should be a lower 
number of gas clumps and that, 
because they do not fragment as much, the clumps should be larger.
Future observations of the dust and all gas phases with comparable
resolution in the Milky Way, Magellanic Clouds and elsewhere in
the local universe will reveal how stars form in low abundance
gas locally and thereby inform models of how stars formed in
the similarly-pristine early universe.

{\bf How do dwarf galaxies build up their stellar mass?} \hskip 5pt
As local analogs of low mass halos in the early universe,
the nearby dwarf galaxies offer unique insight into the processes
by which halos build up their stellar populations across cosmic
time. Future deep fields will explore the progenitors of today's brighter
dwarfs during the era of Cosmic Noon 
(\cite[Boylan-Kolchin et al. 2015]{BoylanKolchin15}), yielding
analogs of both the star-forming dwarfs and the
dSphs. Future deep observations will explore a broad range of
environments yielding insight into the relative importance of the
mechanisms that imprint the observed morphological segregation. 
With increasing depth, CMDs of LG dwarfs will drive down the uncertainty on
individual SFHs and allow the evaluation of satellite populations. 
As noted by Andrew Cole in his overview talk,
it is critical to push photometry to below 28th or 29th
magnitude even for dwarfs within 1 Mpc, setting the benchmark
for studies of the LG. Spectroscopic studies of individual
stars spanning the full lifetime will probe the chemical evolution
and dynamics over a galaxy's history (\cite[Kirby et al. 2017]{Kirby17}).
In combination with detailed orbital reconstructions, it will be
possible to correlate important epochs of a galaxy's SFH with
with the corresponding timesteps in its orbital history.

{\bf What do today's dwarfs tell us about reionization?} \hskip 5pt
Current consensus models suggest that reionization was contributed in large
part by the progenitors of today's dwarf galaxies. 
\cite[Boylan-Kolchin et al. (2015)]{BoylanKolchin15} discuss the importance of
the LG dwarfs to understanding the role of dwarfs in reionization and predict
the range of the rest-frame UV luminosity function accessible to current
and future surveys. Future deep fields with HST and JWST will probe deeper 
and deeper, especially through lensing, down the UV luminosity 
function. Of notable importance, Figure 4 of \cite[Boylan-Kolchin et al. 
(2015)]{BoylanKolchin15} demonstrates
that JWST will be able to probe systems of luminosity comparable to the LMC
at redshifts $\sim$ 7; 
similar objects which benefit from gravitational lensing should be contained among
the faintest objects visible in the HST Frontier Fields. At the same time, 
since the progenitors of the lowest mass LG dwarfs will not be visible at 
high redshift, it is of critical importance to understand the relative contributions 
of galaxies of LMC mass and lower by detailed studies of their local analogs. 
As pointed out by \cite[Weisz et al. (2011)]{Weisz11}, 
deep studies of the LG dwarfs will set much more robust 
constraints on the nature and state of their progenitors at early times. 
And, the detailed study of massive stars in local low metallicity dIs 
will provide a better understanding of massive star formation
in the early universe and how the processes associated with the
evolution of massive stars leaves an imprint on the evolution of
their host dwarf galaxies. 

{\bf Is the Local Group representative?} \hskip 5pt
Detailed observations of the stars, gas and dust in LG dwarfs drive
constraints on numerical simulations and models of galaxy formation that
explore the low mass regime. However, we should ask whether the LG 
really serves as a fair laboratory into galaxy assembly and evolution 
given its relatively, but not entirely, quiescent state. 
As mentioned previously, work on the satellites of M31 
(\cite[Martin et al. 2017]{Martin17}) are
beginning to test our understanding of MW dwarfs in comparison, but the
picture is not yet clear. 
\cite[Weisz et al. (2011)]{Weisz11} have explored the SFHs and morphological 
segregation of LG galaxies
with those of the more distant and widely distributed ANGST sample. 
Within the limitations
of the data available to that study, the LG dwarfs appear to be
fairly representative of other nearby systems though studies of comparable 
depth over considerably larger samples will undoubtedly provide insights 
beyond today's cursory picture. 
As Andrew Cole has reminded us in this volume, the uncertainty in the SFH is a 
strong function of limiting depth, so that comparison is really only fair if 
it is made over the same luminosity range. 

The next generation of telescopes on the ground and in space will offer 
tremendous new capabilities that promise the exploration of other local
volumes for comparison with the LG. Observations are just beginning to
probe nearby groups with sufficient depth. 
For example, \cite[Danieli et al. (2017)]{Danieli17} derive the luminosity 
function of the
M101 group and find that it is flatter at faint magnitudes than that of the
LG. Furthermore, they find an apparent lack of intermediate mass galaxies 
in the M101 group. However,
kinematic measurements are are critical to determining whether there is really
a ``too big to fail'' challenge in the M101 group.  

Intriguingly, \cite[D'Souza \& Bell (2018)]{DSouza18}
have suggested that M32 results from the merger 2 Gyr ago of M31 with a large
galaxy of stellar mass $\sim$2.5 $\times$ 10$^{10}$ \msun, the third 
largest galaxy in the LG. Hence, we need to understand better
the assembly history of the LG in order to ascertain how typical the
present time is.  Only then will we know how representative
the LG is. Given the current uncertainties, we should 
keep an open mind and not rely on the LG too much for assumptions about 
how galaxies across
the universe and cosmic history might behave.

Future observations of dwarf galaxies will provide critical tests of cosmological 
and astrophysical theories and simulations particularly through
the interplay of their luminous and dark components and will refine
the constraints placed by the tiniest galaxies on our understanding 
of galaxy formation and evolution over cosmic time. Given the promised 
capabilities of the next generation of instruments and facilities, 
dwarf galaxy research offers a most promising prospectus for 
the enterprise of astronomical discovery.

Acknowledgements: I want to thank Kristy McQuinn, Sabrina Stierwalt and
Romeel Dav\'{e} for organizing this terrific symposium and all of its participants
for an exciting and stimulating week. I am 
grateful for the support of NSF/AST-1714828 and the Brinson Foundation.

\end{document}